\documentstyle[12pt]{article}

\begin{document}

\title{Astrophysical Polarimetric Signature Against TeV Fundamental Planck Scale}
\author{Yu.\ N.\ Gnedin\\
Central Astronomical Observatory at Pulkovo,
                   Sankt-Petersburg, Russia}

\maketitle

\begin{abstract}
      I present the analysis of data of astrophysical polarimetric
observations that gives the signature of the fundamental extra
dimension Planck scale magnitude essentially higher than $~1TeV$.
Magnetic conversion of photons into the fundamental particles
(scalars, gravitons) is the probable mechanism that can produce 
noticeable amount of polarization of optical radiation of astrophysical
objects, especially, of distant extragalactic sources. The results of
magnetic conversion process of optical light of extragalactic sources
are presented for a number of situations including: (a) intergalactic
magnetic field, (b) galaxy cluster magnetic field, (c) magnetic
conversion in the typical galaxy magnetic field, (d) magnetic
conversion of CMB radiation.
\end{abstract}

\section{Introduction}

         There is now a very popular idea of the existence of additional
dimensions beyond the four ones that we explore in our every day life. A
possibility that the Universe has additional compactified spatial
dimensions has been long discussed (see the classical papers by Kaluza
(1921) and Klein (1926)). This idea has been successfully developed by
modern string theory (see the review by Green et al. (1987)). Recently
this idea has been combined with Grand Unification Theory of
Particle Physics.

         The most outstanding problem in modern physics is to explain
the extraordinary difference between the electroweak scale $ M_{EW} =
10^{3}GeV$ and the four-dimensional Planck scale $M_{p}=10^{19}$ Gev.
The various scenarios have been proposed. For example, the Standard
Model (SM) of Particle Physics is localized on a three dimensional
brane in a higher dimensional space with large compactified
space-like extra dimensions. In a model suggested by Arkani-Hamed et
al.(1998,1999), the matter is confined to a 3-brane while the gravity
propagates in extra dimensions of a sub-millimeter size. These new
dimensions should be sufficiently compact so as to escape trivial
detection.

         The main exciting consequence of these new theories is
possibility that the Planck, string and Grand Unification scales can
all be significantly lower than it was previously thought, perhaps as
low as few TeV (see, far example, Antoniadis (1990), Lykken(1996),
Shiu and Tye (1998), Antoniadis and Baches (1999), Kubyshkin
(2001),etc). Also in theories with standard model of gauge bosons
propagating in $TeV^{-1}$-size extra dimensions, their Kaluza-Klein
states can interact with the rest of SM particles confined to the
3-brane.

Many physicists try to look for possible signals for this
interaction in the present high-energy collider data, and estimate
sensitivity that can be reached by the next generation of collider
experiments (see recent publications by Bordag et al. (2000), Lykken
and Nandi (2000), Cheung and Landsberg (2001), McMullen and Nandi (2001)
and Muck et al. (2001)).

        "n" extra dimensions are compactified at the scale $R^{-1}$,
where the size $R$ is related to the four-dimensional Planck scale
$M_{p}$ via the relation:
\begin{equation}
M_p^2=M_s^{n+2} R^n,
\end{equation}
where the new scale $ M_{s}$ is the fundamental $ (4+n)$- dimensional Planck
scale which appears of the same order as the string scale.

        More recent studies have shown that there could be possible
scenarios of stringly nature where $ 1/R $ and $M_s$ may be lowered
independently of $M_{p}$ by several of many orders of magnitude. In
particular, the scenario is now very popular with radical possibility
that $M_{s}$ of order TeV. In such a way $M_{s}$ represents the only
fundamental scale in the Universe at which unification of all forces
of nature occurs. Arkani-Hamed et al.(1998) have developed this
scenario and made the conclusion that the compactification radius is
related to the higher dimensional gravitational interactions lies in
the sub-mm range, i.e. $1/R < 10^{-3} eV$. They hope that
Cavendish-type experiments may potentially test the model by observing
deviations from Newron's law at such small distances. Their model can
be also tested by high-energy collider experiments.

        One of the direct consequence of extra dimension universe is
possible variation of the fine structure constant, especially, at high
redshifts. Dirac (1938) was among the first to suggest that
fundamental constants, such as the fine structure constant $\alpha =
e^{2}/\hbar c$ could vary with time. The interest in varying constant
theories has recently risen with the increased popularity of above
mentioned models with extra dimensions (see, for instance, Damour and
Dyson(1996), S. Caroll(1998), Varshalovich et al.(2000), Murphy et
al.(2001a,b), Webb et al.(2001), Barrow et al.(2001)).

        The various  astrophysical effects can lead to constraints on
the effective fundamental scale $M_{s}$ (see the excellent last review
by Kubyshkin (2001), Table 3 from this review and references therein).

        The goal of this paper is to include in astrophysical effects 
that can give essential bounds on $M_{s}$ the recent polarimetric
observations of extragalactic objects with high redshifts (active
galaxy nuclei (AGNs), quasars (QSOs), radiogalaxies, galaxy
clusters, etc). The central idea is to investigate the new process that
can produce a noticeable amount of polarized light into galaxy and
galaxy cluster's magnetic field, and also into intergalactic magnetic
field. I mean the process of the magnetic conversion of radiation into
(pseudo) scalars and gravitons. This process has been considered as
the real mechanism for production of polarized light in astrophysics
by Raffelt and Stodolsky (1988), Harari and Sikivie (1992), Gnedin and
Krasnikov (1992), Gnedin (1994). Raffelt and Stodolsky have considered
this process and for gravitons. Let us mentioned that the process of
magnetic conversion was specially explored due to searches for axions.

        I shall show below that polarimetry of extragalactic objects
gives more strong bounds on the fundamental scale $M_{s}$ than this is to
be expected from collider experiments and other astrophysical effects
(cooling of the Universe, SN1987 cooling and cosmic diffuse gamma
radiation).

\section{Magnetic Conversion of Photons into Fundamental Particles}

        Grand Unification theory (GUT) requires the existence of
coupling between photons and fundamental particles. This coupling is
determined by Lagrangian term (for scalars):
\begin{equation}
-\frac{1}{M_s}\phi F^{\mu\nu}F_{\mu\nu},
\end{equation}
where $F$ is the tensor of electromagnetic field and $\phi$ is a scalar
field.

       The theory gives the following expression for probability of
conversion of definitely polarized photons $W_{||}$ into scalar particles
(Raffelt and Stodolsky (1988), Gnedin (1994)):
\begin{equation}
W_{||} = \frac{L_p^2}{L_B^2+L_p^2}\sin^2\left(\frac{1}{2}
\frac{BL_{coh}}{M_s}\sqrt{1+L_B^2/L_p^2}\right),
\end{equation}
where $B$ is the magnetic field strength, $L_{coh}$ is the coherence
length of magnetic field, $L_{B} = 2\pi M_{s}/B$ and $L_{p} =
2\pi\omega/ \omega_{p}^{2}$ are the oscillation lengths of magnetic
conversion into vacuum magnetic field and into plasma, respectively.
Only one polarization state for which the electric vector lies into
the plane containing the magnetic field and line of sight directions
is transformed. Here and below the symbol $B$ means really the
projection of the vector $B$ on the this plane.

        The Eq.(3) is valid only if the condition
$L_{B},L_{p}<2\pi\omega/m_{\phi}$ takes place, where $m_{\phi}$ is the mass of
a scalar. Therefore, our consideration is restricted only by low mass
and massless scalars or gravitons.

        For the case of vacuum, i.e. when $L_{p}\gg L_{B}$ Eq.(3) is very
simplified and takes a form:
\begin{equation}
W_{||} = \sin^2\left(\frac{1}{2}\frac{BL_{coh}}{M_s}\right) \approx
\frac{B^2L^2_{coh}}{4M_s^2}
\end{equation}
if the condition takes $BL_{coh}\ll M_{s}$.
  
        The degree of linear polarization $p_{l}$ can be easily found
by  
\begin{equation}
P_l = \frac{I_\perp-I_{||}(1-W_{||})}{I_\perp+I_{||}(1-W_{||})}
\approx W_{||}/2
\end{equation}
if one has deal with nonpolarized light, i.e. $I_{||} = I_{\perp} =
I_{0}/2$ and $W_{||}\ll1$.

        Now the main problem consists in the estimation of the
magnitudes of $B$ and $L_{coh}$ for real astrophysical conditions.

\section{Magnetic Field Strength and the Coherent Length in the
Universe}

       Magnetic field play an important role in practically all
astrophysical phenomena. There are some of the reviews and papers
concerning to the origin and possible effects of magnetic fields in the
Universe and also to the current status of the art of observations of
cosmic magnetic fields (see, for example, Kronberg(1994) Grasso and
Rubinstein(2001), Carilli and Taylor (2001), Dolgov (2001), Gnedin et
al.(2000), Furlanetto and Loeb (2001)).

       Let us start with situation of magnetic fields in galaxies. The
interstellar magnetic field in the Milky Way has been determined by
several methods which gave valuable information about the amplitude
and spatial structure of the field. The average field strength is $3-4
microG$. Such a strength corresponds to an approximate energy
equipartition between the magnetic field, the cosmic rays confined in
the Galaxy, and the small-scale turbulent motion.

      Observations on a large number of Abel clusters, some of which
have a measured X-ray emission, have given valuable information on
fields in cluster of galaxies (Kim et al.1991). Magnetic field
strength in the inter cluster medium (ICM) can be quite well estimate
by the phenomenological relation (Grasso and Rubinstein 2001):
\begin{equation}
B_{ICM} \approx 2\mu G\left(\frac{L_{coh}}{10 kpc}\right)^{-1/2}
h_{50}^{-1}.
\end{equation}

      Typical values of $L_{coh}$ are 10-100 kpc which correspond to
field magnitudes of $10 - 1 \mu G$. For example, the case of the Coma
Cluster a core magnetic field strength reaches $B\approx8.3h_{100}^{1/2}\mu G$ at
scales of about 1 kpc. An exciting example of clusters with a strong
magnetic field is the Hydra A cluster for which the Rotation Measure
(RM) implies a 6 microGauss field over 100 kpc superimposed with a
tangled field of strength 30 microGauss (Taylor and Perley,1993).
The high-resolution images of radio sources embedded in galaxy
clusters show evidence of strong magnetic fields in the cluster
regions, and also in the regions of cool fronts and cool fluxes
(Carilli and Taylor, 2001). The typical central field strength
is approximately 10
-30 microGauss with the peak values as large as  70 micro Gauss. 

       Furlanetto and Loeb (2001) gave an estimation of the magnetic
field strength in the diffuse intergalactic medium (IGM) assuming flux
conservation for out flows from QSOs that inevitably pollute IGM. They
obtained $B_{IGM}\sim10^{-9}G$ with the coherence length ~ 1 Mpc. The
observational constraints on an IGM field imply more soft bounds,
requiring only that $B_{IGM} < 106{-8}(L_{coh}/Mpc)^{-1/2}G$ with use
of the currently popular $\Lambda CDM$ model.

      The last exciting result has been recently obtained by
Hutsemekers and Lamy (2000), who discovered the existence of coherent
orientations of QSO polarization vectors on cosmological scales.
Considering a sample of 170 optically polarized QSOs with accurate
polarization measurements they found that QSO polarization vectors are
not randomly oriented on the sky as naturally expected. They claim
that these observations give an evidence for the presence of
correlations, probably, IGM magnetic field on spatial scales
$L_{coh}\sim10^{3}h^{-1}$ Mpc at redshifts $z \approx1-2$.

       Now let us start with estimation of the fundamental extra
dimension Planck scale $M_{s}$ with use of recent polarimetric data of
observations of QSOs and AGNs in optical range.

            4. Estimation of Fundamental Extra Dimension Planck Scale
from Optical Polarimetric Data.

\subsection{Magnetic Photon Conversion in the IGM}

        We shall make our estimations using approximation by Furlanetto
and Loeb (2001) accepting the dependence of IGM magnetic field
strength on coherence length in a form
\begin{equation}
B \equiv B_{ICM} = 10^{-9}\left(L_{coh}/1 Mpc\right)^{-1/2} G.
\end{equation}
  The IGM electron density is
\begin{equation}
n_e = \Omega_bh^2\times 10^{-5}(1+z)^3 cm^{-3} \approx 2\times
10^{-7}(1+z)^3 cm^{-3}.
\end{equation}

 The the oscillations lengths are:
\[
L_p = \frac{2\pi\omega(1+z)}{\omega_p^2}\approx
2\times10^{29}\left(\frac{\omega}{3 eV}\right)
\frac{1}{(1+z)^2} eV^{-1},
\]
\begin{equation}
L_B = \frac{2\pi M_s}{B} = 10^{23} \left(\frac{10^{-9} G}{B}\right)
\left(\frac{M_s}{1 TeV}\right) eV^{-1},
\end{equation}
where $\omega_p$ is the plasma frequency.

  The commonly accepted system units $\hbar=c=1$ is here used. Eq.(9)
means that $L_{B} < L_{p}$ if $M_{s}<10^{5}$ TeV. We consider the case
of high redshift objects with $z\leq2$. For the extragalactic objects
with $z\leq1$ the condition $M_{s} \leq 10^{6}$ TeV requires that
$L_{B}\leq L_{p}$.

         Now let calculate the value of extra dimension Planck scale
directly from Eq.(3). The polarization level at 0.01 is quite well
consistent to observable data (see review Koratkar and Blaes (1999)
and references therein and Hutsemekers and Lamy (2000)). The coherence
length $L_{coh}\sim 1 Mpc$ appears to be larger that the oscillations
lengthes $L_{p}$ and $L_{B}$ but $L_{p} < L_{B}$. Then
\begin{equation}
P_l \approx 0.01 \approx \frac{L_p^2}{L_B^2}
\sin^2\left(\frac{1}{2}\frac{BL_{coh}}{M_s}
\sqrt{1+L_B^2/L_p^2}\right).
\end{equation}
From ratio $L_{p}^{2}/L_{B}^{2} \approx 10^{-2}$ it follows:
\begin{equation}
M_s \approx 10^6 TeV \left(\frac{B}{10^{-9}G}\right)
\left(\frac{\omega}{3 eV}\right)
\end{equation}
for $z\approx2$.

        Let us put a question what IGM magnetic field strength
corresponds to $M_{s}\approx 1TeV$. One needs to require two conditions:
$L_{B}\gg L_{coh}\gg L_{p}$ and $L_{coh}/L_{B}\sim 0.1$. The last condition is
required that polarization $P_{l}$ exists at the observable level of 1 percent..
One can get from these conditions the relation $2\pi M_{s}/B > 10 Mpc$
and $B < 4 10^{-18}G$. This IGM magnetic field seems to be very small
and inconsistent with recent observable polarimetric data obtained by
Hutsemekers and Lamy (2000).

\subsection{Magnetic Photon Conversion in Fields of Galaxy Clusters}

     Let us estimate the magnetic photon conversion rate in cluster
magnetic fields using the data from most recent review by Carilli and
Taylor (2001) (see Table 1 from this review). The typical magnetic
field strength in the cluster is to be $B\approx10\mu G$ and the coherence
length (cell) size is ~ 10 kpc. The most commonly accepted value of
the central density of a cluster is $n_{0} \approx10^{-3} - 10^{2}
cm^{-3}$. For this situation the parameters of magnetic conversion
process are:
\[
L_B \approx 10^{20}\left(\frac{M_s}{10TeV}\right)
\left(\frac{10\mu G}{B}\right) eV^{-1},
\]
\begin{equation}
L_p \approx 6\times10^{23}\left(\frac{\omega}{3eV}\right)
\left(\frac{10^{-2} cm^{-3}}{n_0}\right) eV^{-1}.
\end{equation}

Because of $L_{B}\ll L_{p}\ll L_{coh}$ the strong oscillations  in
magnetic conversion process take place and one can expect the
polarization degree for embedded or background optical source at the
high level $p_{l} \sim 100\%$, that contradicts really to observable data. 

      For $n_{0} \approx 10^{-3} cm^{-3}$ and the observable polarization
level $\approx1\%$ one can estimate the value of the effective Planck scale as
$M_{s} \approx 10^{7} TeV$.

\subsection{Magnetic Photon Conversion in Galaxies}

     The basic parameters of magnetic conversion into typical galaxy
are:
\[
L_B \approx 3\times10^{19}\left(\frac{M_s}{1TeV}\right)
\left(\frac{3\times10^{-6} G}{B}\right) eV^{-1},
\]
\begin{equation}
L_p \approx 2\times10^{21}\left(\frac{\omega}{3eV}\right)
\left(\frac{1 cm^{-3}}{n_e}\right) eV^{-1}.
\end{equation}

Here $n_{e}\approx1 cm^{-3}$ is the average electron density in a typical
galaxy. $M_{s}\sim 1 TeV$ is excluded because this case gives such high
polarization value$ \approx 100\%$, more higher compare to observed
interstellar polarization that has a level at some percents.

       If one wants to estimate the contribution of magnetic
conversion process, this process needs to provide the level of
polarization at least not higher than the interstellar polarization.

       For the case of strong coupling photons to scalars when
$L_{p}, L_{B} < L_{coh}$, the condition $L_{p} < L_{B}$ is required
for obtaining  the polarization comparable to interstellar one from
the nearest stars located at the distance $L \approx 100 pc$, and then
$M_{s} \approx 10^{3}TeV$. 
 
       The case of weak coupling is accomplished if only $L_{p} >
L_{B} > L_{coh}$. In this case magnetic conversion polarization is
determined by Eq.(4) with replace of $L_{coh}$ on $L$. If, for
example, $n_{e}\approx10^{-4}$ (low density regions of a galaxy), the
Eq.(4) provides by comparison with the interstellar polarization
$M_{s}\geq 3*10^{6} TeV$.

        It is interesting to notice that there are polarimetric
observations that show the violation of interstellar polarization
wavelength curve for a number of stars from the famous Serkowskii law,
the observable polarization being increased in near UV range (this
fact is in favor of magnetic conversion mechanism). Of course, there
exists another explanation of this discrepancy that looks not so
exotic (see Martin et al.1999).

\section{Magnetic Photon Conversion as a Cause of Dependence of Fraction
Polarized QSOs on Redshift}

    Impey et al.(1991), Wills et al.(1992) and Carilli et al.(2000)
have shown that a fraction of extragalactic sources (AGNs, QSOs,
superluminous IR galaxies) with noticeable optical polarization
magnitude is really decreasing with increase of redshift magnitude.
There are probably two ways for explanation of this phenomenon: this
is an intrinsic cause due , for example, to Faraday depolarization in
accreting disks around supermassive black holes (see Gnedin and
Silant'ev (1997, 2001)) and the external cause due to depolarization
in extended environment and IGM.

     We are here considering the second possibility suggesting that
depolarization is produced by birefrigent effect of magnetic photon
conversion into low mass scalars or gravitons (see Raffelt and
Stodolsky (1988), Gnedin and Krasnikov (1992)), the rotating angle
being not dependent on wavelength.

      In the case of strong coupling when $L_{B}\ll L_{p}\ll L_{coh}$ the
optical radiation of distant extragalactic sources should be
completely depolarized, but it is not this case. In the case of weak
coupling when the rotation angle is determined by the following
expression (Raffelt and Stodolsky (1988)):
\begin{equation}
\Theta = \frac{1}{8}\frac{B^2L_{coh}^2}{M_s^2}\approx 1,
\end{equation}
there is noncomplete depolarization that corresponds better to
observable data.

      Then we obtain the following estimation of the fundamental extra
dimension Planck scale:
\begin{equation}
M_s \geq 3\times10^6 TeV \left(\frac{B}{10^{-9}G}\right)
\left(\frac{L_{coh}}{1 Mpc}\right).
\end{equation}
Of course the Eq.(15) does not exclude the case $M_{s} \approx 1 TeV$ if the
magnetic field of IGM takes magnitude $~ 10^{-15}G$ for $z \approx 1-2$, but
this fact seems quite unreal.

\section{Magnetic Photon Conversion Process and CMB}

      The question is arising how the magnetic conversion of photons to
fundamental particles affects on the CMB, in particular, producing its
polarization. We are going to discuss this problem in detail in the separate paper.
 Here we would like only roughly to estimate the effect. For the CMB the parameters 
of magnetic conversion are:
\[
L_B \approx 10^{23}\left(\frac{M_s}{1TeV}\right)
\left(\frac{10^{-9} G}{B}\right) eV^{-1},
\]
\begin{equation}
L_p \approx 10^{26}(1+z)^{-2}
eV^{-1}.
\end{equation}
and $L_{coh}>L_{p}>L_{B}$ that means that strong coupling is acting in
this case. The degree of polarization of CMB would be high at the
level close to $ 100\%$. But more weak magnetic fields requires more
high magnitude of the fundamental Planck scale. It means that the
validity of Eq.(16) requires the strong bounds on IGM magnetic field
and the fundamental Planck scale, namely,
\begin{equation}
\left(\frac{10^{-9}G}{B}\right)\left(\frac{M_s}{1TeV}\right)
\leq 10^3(1+z)^{-2}.
\end{equation}
This result provides that if $M_{s}\geq 1TeV$ one can expect the strong
polarization of CMB at angular scales $ <~ 1 arcmin$.

\section{Conclusions}

       Though our results does not completely exclude the fundamental
Planck scale magnitude $M_{s} \sim1TeV$ , the available polarimetric
data, especially, for extragalactic sources makes the effective Planck
scale magnitude $M_{s} \gg 1 TeV$ more probable.

       Our results can be presented by the following Table 1.

\begin{table}
\caption{Effect of Magnetic Conversion on Polarimetric Data}
\renewcommand{\arraystretch}{1.5}
\begin{tabular}{l|c|c|c}
\hline
    Object of Polarimetric Observa-  &
  Magnetic field & Coherence length  &
 Planck scale \\
 tions in the optical range &  $B$ & $L_{coh}$ & $M_s$ \\
\hline
Intergalactic Medium & $B\leq10^{-9}$G & $L_{coh}\sim1$Mpc  &  $M_{s}\geq10^{6}$TeV \\
Galaxy Clusters      & $B\sim10^{-5}$G & $L_{coh}\sim10$kpc &  $M_{s}\geq10^{7}$TeV \\
Typical Galaxy       & $B\sim3\times10^{-6}$G & $L_{coh}\sim100$pc &
    $M_{s}\geq10^{3}$TeV \\
Fraction of Polarized QSOs   &
                $B\leq10^{-9}$G & $L_{coh}~1$Mpc & $M_{s}\leq10^{7}$TeV \\
and AGNs  & & & \\
   Cosmic Microwave Background  &          $B\leq10^{-9}$G & $L_{coh}~1Mpc$
& $M_{s}\leq10^{3}$ \\
\end{tabular}
\end{table}                                          

      At last, let us remind that we considered only the process of
photon magnetic conversion into low mass and massless scalars and
gravitons. The case of massive scalars and gravitons requires special
consideration.

      This work is supported by Grants of RFFI, No.99-02-16366 and of
the Program "Nonstationary Phenomena in Astrophysics" and by the
Program "INTEGRATION", No.K032. I would like to express many thanks to
people from the CASA and JILA, Boulder, Colorado for their warm
hospitality.

\begin{center}
\Large\bf References
\end{center}
\def\ref{\par\noindent\hangindent=3pc\hangafter=1}

\ref Antoniadis I. 1990, Phys.Lett., B246,377.
\ref Antoniadis I., Bashes C. 1999, Phys.Lett.,B450,83.
\ref Arkani-Hamed N., Dimopoulos S.,Dvali G. 1999,Phys.Rev.,D59,086004.
\ref Arkani-Hamed N.,Dimopoulos S.,Dvali G., 1998,Phys.Lett.,B429,263.
\ref Barrow J.D.,Sandvik H.B.,Magueijo J. 2001, astro-ph/0109414.
\ref Bordag M.,Geyer B.,Klimnitskaya C.L.,Mostepanenko V.M.,2000,Phys.Rev.,
                                                       D62, 011701.
\ref Carilli C.L. et al. 2000, astro-ph/0008380.
\ref Carilli C.L., Taylor G.B. 2001, astro-ph/0110655.
\ref Caroll S. 1998, Phys.Rev.Lett.,81,3067.
\ref Cheung K., Landsberg G. 2001, hep-ph/0110346.
\ref Damour T., Dyson F. 1996, Nucl.Phys.,B480,37.
\ref Dirac P.A.M. 1938, Proc.Roy.Soc.Lond.,A165, 199.
\ref Dolgov A.D. 2001, astro-ph/0110293.
\ref Furlanetto S.R., Loeb A. 2001, astro-ph/0110090.
\ref Gnedin N.Y., Ferrara A.,Zweibel E.G. 2000, Ap.J., 539, 505.
\ref Gnedin Yu.N. 1994, Astron.Astrophys.Trans., 5, 163.
\ref Gnedin Yu.N., Krasnikov S.V. 1992, Sov.Phys.JETP, 75, 933.
\ref Gnedin Yu.N., Silant'ev N.A. 1997, Basic Mechanisms of Light
\ref Polarization in Cosmic Media, Hartwood Academic Publ., Amsterdam, p.30.
\ref Gnedin Yu.N., Silant'ev N.A. 2001, Pis'ma Astron.Zh., to be published.
\ref Green M.B, Schwartz J.H., Witten E. 1987, Superstring Theory, Cambridge,
Cambridge Univ.Press.
\ref Grasso D., Rubinstein H.R. 2001, Phys.Rep., 348, 163.
\ref Harari D., Sikivie P. 1992, Phys.Lett., B289, 67.
\ref Hutsemekers D., Lamy H. 2000, astro-ph/0012182.
\ref Impey C.D., Lawrence C.R., Tapia S. 1991, Ap.J., 375, 46.
\ref Kaluza T. 1921, Preuss.Akad.Wiss., Berlin, p.966.
\ref Kim K.T., Kronberg P.P., Tribble P.G. 1991, Ap.J., 379, 80.
\ref Klein O. 1926, Z.Phys., 37, 895.
\ref Koratkar A., Blaes O. 1999, PASP, 111, 1.
\ref Kronberg P.P. 1994,Rep.Prog.Phys., 57, 325.
\ref Kubyshkin Yu.A. 2001, hep-ph/0111027.
\ref Lykken J. 1996, Phys.Rev., D54,3693.
\ref Lykken J., Nandi S. 2000, Phys.Lett., B485, 224.
\ref Martin P.G., Clayton G.C.,Wolff M.J. 1999, Ap.J., 510, 905.
\ref McMullen C.D., Nandi S. 2001, hep-ph/0110275.
\ref Murphy et al. 2001a, MNRAS, 327, 1237; 2001b, MNRAS, 327, 1244.
\ref Muck A., Pilaftsis A., Ruckl R. 2001, hep-ph/0110391.
\ref Raffelt G., Stodolsky L. 1988, Phys.Rev., D37, 1237.
\ref Shiu G.,Tye S. 1988, Phys.Rev., D58, 106007.
\ref Taylor G.B., Perley R.A. 1993,Ap.J., 416, 554.
\ref Varshalovich D.A., Potekhin A.Y., Ivanchik A.V.
2000,AIP.Conf.Proc.(Melville), 506, 503.
\ref Webb et al. 2001, Phys.Rev.Lett., 87, 091301.
\ref Wills B.J., Wills D., Breger M., Antonucci R.R.J., Barvians R.
1992, Ap.J., 398, 454.
 
\end{document}